\documentclass[conference]{IEEEtran}
\IEEEoverridecommandlockouts

\usepackage[T1]{fontenc}
\usepackage{cite}
\usepackage{amsmath,amssymb,amsfonts}
\usepackage{graphicx}
\usepackage{textcomp}
\usepackage{booktabs}
\usepackage{url}
\usepackage{xurl}
\usepackage{microtype}
\usepackage{float}

\begin{document}

\title{Impact of Wind-Solar-Battery Growth on Electricity Prices: Reserve Scarcity and Transmission Congestion
 
}

\author{\IEEEauthorblockN{Haoxiang Wan\IEEEauthorrefmark{1}, Ivan Castillo\IEEEauthorrefmark{2}, and Xingpeng Li\IEEEauthorrefmark{1}}
\IEEEauthorblockA{\IEEEauthorrefmark{1}\textit{Dept. of Electrical and Computer Engineering, University of Houston}, Houston, TX, USA \\
hwan6@cougarnet.uh.edu, xli83@central.uh.edu}
\IEEEauthorblockA{\IEEEauthorrefmark{2}\textit{Enterprise, Innovation and Growth (EIG), Applied Digital Sciences (ADS)} \\
\textit{The Dow Chemical Company}, Lake Jackson, TX, USA \\
castillo2@dow.com}}
\maketitle

\begin{abstract}
This paper asks whether sequential wind, solar, and battery storage additions that
eliminate unserved energy also remove modeled reserve scarcity and transmission
congestion. Thirty-one cumulative portfolio states are evaluated on a modified
IEEE 24-bus system for a constructed stressed day using a 24-hour day-ahead
unit commitment model with network constraints, co-optimized energy and upward
reserve, chronological battery storage, and an approximation of the Electric
Reliability Council of Texas operating reserve demand curve. The thermal
baseline has 106 MWh of unserved energy, whereas 450 MW of wind is sufficient
to serve all demand. However, with 1{,}500 MW of wind, the hourly
modeled reserve scarcity price still reaches \$5{,}000/MWh because wind output declines
after the system peak. Adding 1{,}800 MW of solar reduces the load weighted
daily average modeled reserve scarcity price from \$1{,}136.2/MWh to \$101.2/MWh and
confines modeled reserve scarcity to three evening hours. Adding 250 MW/1{,}000 MWh of battery storage
further lowers the average to \$4.3/MWh and reduces wind curtailment from
631 MWh to 306 MWh. Transmission congestion nevertheless remains: all residual
wind curtailment occurs at bus 21, wind capture prices at buses 21 and 23 differ
by \$31.8/MWh, and identical battery storage plants yield normalized net energy market
revenues of \$201/(MW$\cdot$day) and \$121/(MW$\cdot$day) at buses 19 and 21, respectively. These
results show that unserved energy, modeled reserve scarcity, and locational
congestion effects measure distinct operating conditions and should be
evaluated separately but interpreted together.
\end{abstract}


\section{Introduction}
\label{sec:introduction}

Adding capacity can remove an energy shortfall without resolving the next
operating limitation. A power system may serve all demand during the peak while
upward reserve remains scarce later in the day. Even when energy and reserve are
sufficient in aggregate, transmission limits can curtail renewable output at
one bus and separate resource values across the network. Evaluating a resource
addition therefore requires attention to how much capacity is added, when it is
available, and where it connects.

This distinction is increasingly important as electricity demand grows and new
resources face long interconnection delays. The 2025 Long-Term Reliability
Assessment of the North American Electric Reliability Corporation identifies
resource adequacy challenges in 13 of 23 assessment areas and notes that additions expected in the near term are concentrated in solar
photovoltaics and batteries \cite{nerc}. Lawrence Berkeley National Laboratory reports more than
2{,}060~GW of generation and storage in U.S. interconnection queues and a
median interval exceeding five years from interconnection request to commercial
operation for projects completed in 2025 \cite{queued}. These conditions make
it necessary to distinguish energy adequacy at the peak, reserve availability in
critical hours, and deliverability through the transmission network.

Previous research provides established concepts for these effects. Large solar
additions can shift the hours of highest net load toward the evening \cite{duck},
while transmission constraints and spatial flexibility can materially affect
renewable curtailment and operating outcomes \cite{bird,rameshcurt,3,2, wan2026ias}.
Operational detail can alter the resource mix selected by planning models
\cite{palmintier}. Market value, capture price, and value factor metrics
characterize the economic value received by variable generation
\cite{mills12,hirth,mills15}; capacity credit methods quantify the contribution
of variable generation and storage to resource adequacy
\cite{keane,wang,denholm}; operating reserve demand curves connect reserve
availability with scarcity pricing \cite{hogan,ercot}; and storage studies show
that temporal price variation and network location both affect storage value
\cite{sioshansi,pandzic,rameshstor}. These outcomes are often examined
separately, making their joint evolution under a common operating data set less
clear.

This study follows a prescribed wind--solar--battery storage sequence under fixed load,
renewable availability, outage, transmission, and operating assumptions. It
asks when unserved energy disappears, how solar and battery storage change the timing
and magnitude of modeled reserve scarcity, and how transmission constraints
affect curtailment, capture prices, and battery storage energy market revenue after all
demand is served. The sequence is not an optimized expansion plan; it is a
controlled comparison in which one resource build-out level changes at a time.

We apply this design to a modified IEEE 24-bus reliability test system based on
\cite{rts96, ramesh2021cdr}. Each of the 31 portfolio states is solved with the same 24-hour
day-ahead unit commitment model. The model co-optimizes energy and upward
reserve, represents chronological battery storage and a lossless direct current (DC)
power flow network, and includes a piecewise constant approximation of the
Electric Reliability Council of Texas (ERCOT) operating reserve demand curve
(ORDC). After each mixed-integer solution, the binary decisions are fixed and
the resulting convex quadratic program is solved again to obtain locational
marginal prices and constraint dual variables.

The analysis makes three contributions. First, it isolates the operating response to sequential wind, solar, and battery storage additions on one constructed stressed day. Second, it reports systemwide outcomes, namely unserved energy, modeled reserve scarcity prices, and congestion rent, together with locational outcomes, namely wind curtailment, capture prices, and normalized net energy market revenue, within one internally consistent model. Third, it distinguishes the role of each addition in this case: wind removes the energy shortfall at the peak, solar and battery storage reduce the remaining evening modeled reserve scarcity, and transmission congestion continues to differentiate outcomes across buses.

\section{Study Design}
\label{sec:design}

\begin{figure}[!t]
\centering
\includegraphics[width=\columnwidth]{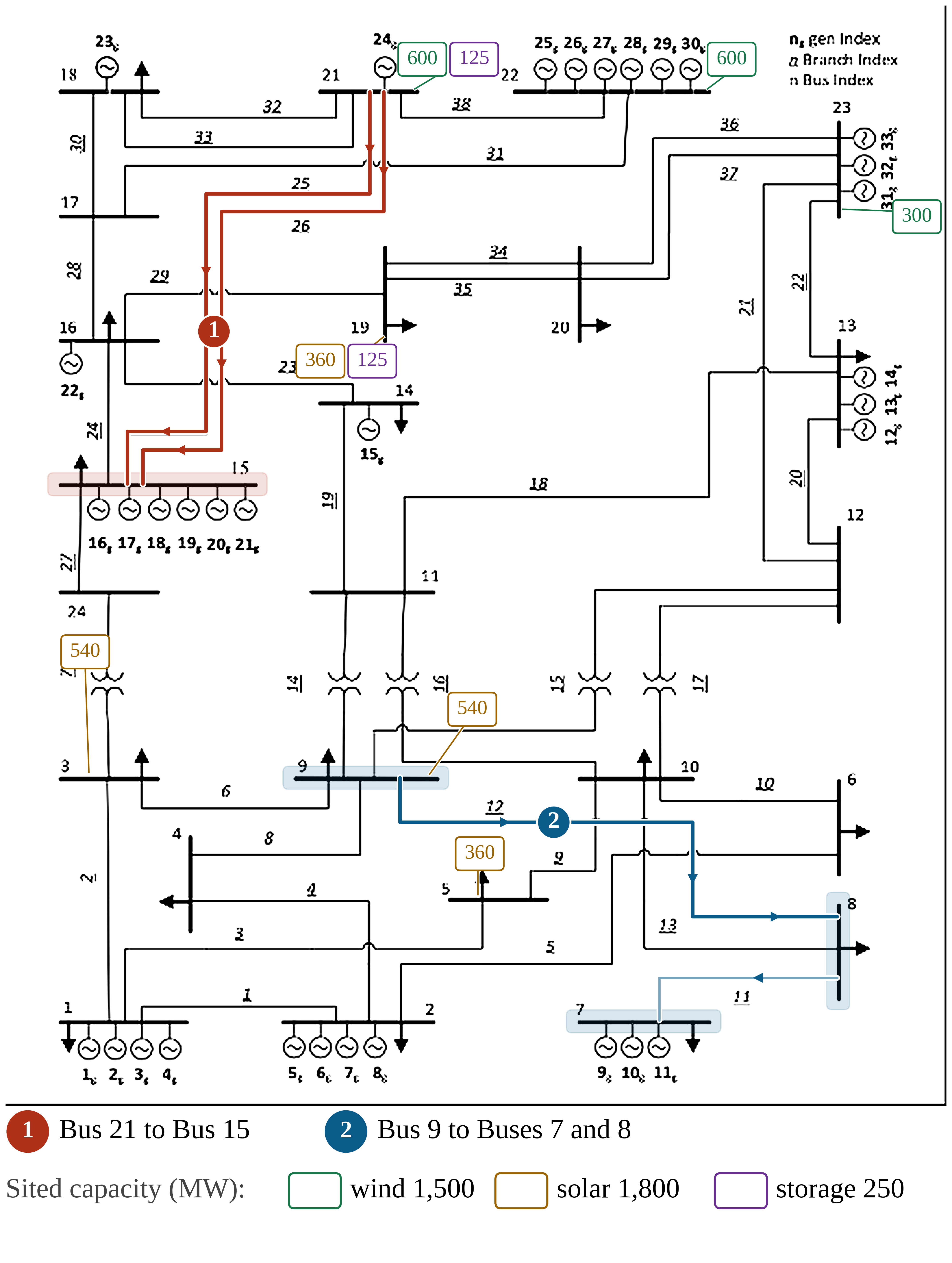}

\vspace{-5mm}
\caption{Modified IEEE 24-bus system, candidate resource locations, and
highlighted delivery interfaces.}
\label{fig:network}
\end{figure}

\subsection{Evaluation Framework}
\label{sec:evaluation}

Each portfolio state is evaluated in the same order. We first determine whether
all demand is served, then examine the timing and magnitude of modeled reserve
scarcity, and finally assess transmission congestion and outcomes at individual
resource locations.

Let $\mathcal B$ denote the set of buses, $\mathcal L$ the set of transmission
branches, and $\mathcal T=\{1,\ldots,24\}$ the set of hourly intervals. Each
interval has duration $\Delta t=1~\mathrm{h}$. The intervals are reported using
hour ending notation, denoted HE1 through HE24. Total unserved energy over the
study day is
\begin{equation}
E^{\mathrm{sh}}
=
\sum_{b\in\mathcal B}\sum_{t\in\mathcal T}
d^{\mathrm{sh}}_{b,t}\Delta t,
\label{eq:unserved_energy}
\end{equation}
where $d^{\mathrm{sh}}_{b,t}$ is unserved load at bus $b$ in interval $t$,
measured in MW. Thus, $E^{\mathrm{sh}}$ is measured in MWh, and
$E^{\mathrm{sh}}=0$ indicates that all demand is served during the study day.

Modeled reserve scarcity is evaluated using the hourly modeled reserve scarcity price
$\pi_t^{\mathrm r}$, defined in Section~\ref{sec:reserve}. We report its hourly
magnitude and the number of intervals with $\pi_t^{\mathrm r}>0$. Let
$D_{b,t}$ denote demand at bus $b$ in interval $t$. For any hourly or nodal
price component $X_{b,t}$, its load weighted daily average, denoted by angle
brackets, is
\begin{equation}
\langle X\rangle
=
\frac{\displaystyle
\sum_{b\in\mathcal B}\sum_{t\in\mathcal T}X_{b,t}D_{b,t}\Delta t}
{\displaystyle
\sum_{b\in\mathcal B}\sum_{t\in\mathcal T}D_{b,t}\Delta t}.
\label{eq:weighted_average}
\end{equation}
Because $\pi_t^{\mathrm r}$ does not vary by bus, its load weighted daily
average $\langle\pi^{\mathrm r}\rangle$ follows from
\eqref{eq:weighted_average} by setting $X_{b,t}=\pi_t^{\mathrm r}$. It equals
the modeled reserve scarcity component of gross load payment divided by total
demand energy.

Transmission effects are evaluated using daily congestion rent and wind
curtailment. Capture prices and normalized net energy market revenue are
reported by location to show how network constraints affect individual
resources.

\subsection{Network and Candidate Resources}
\label{sec:portfolio}

The study uses a modified IEEE 24-bus reliability test system based on
\cite{rts96, ramesh2021cdr} and represents it as one balancing area. The study day
represents a hot summer day with tight supply. Demand peaks at 3{,}100 MW and stays high into the evening. Wind and solar availability
both fall in the evening, so renewable output is low while demand is still
high. A 197 MW unit and a 100 MW unit are out of service for the whole day,
and another 100 MW unit is derated to 50\% of its rating from HE15 to HE21.
These outages reduce available thermal capacity at the peak to 3{,}058 MW.
Thermal capacity alone therefore cannot serve peak demand, and the evening
hours test whether added resources provide enough upward flexibility after
the peak.

The candidate portfolio contains
1{,}500 MW of wind at buses 21, 22, and 23, allocated in nameplate shares of
0.4, 0.4, and 0.2, respectively; 1{,}800 MW of solar at buses 3, 5, 9, and 19,
allocated in shares of 0.3, 0.2, 0.3, and 0.2, respectively; and
250 MW/1{,}000 MWh of battery storage divided equally between buses 19 and 21.
Each battery storage plant therefore has a 125 MW/500 MWh nameplate rating and a 4 h
nameplate duration. Production incentives of \$30/MWh for wind and \$5/MWh for
solar are represented as offer costs of $-\$30/\mathrm{MWh}$ and
$-\$5/\mathrm{MWh}$, respectively.

The superscripts $\mathrm w$, $\mathrm s$, and $\mathrm e$ denote wind, solar,
and battery storage. For each resource type
$x\in\{\mathrm w,\mathrm s,\mathrm e\}$, the build-out level
$\rho^x\in[0,1]$ is the installed fraction of the corresponding candidate
capacity. The same build-out level is applied at every candidate site of a
resource type, preserving the spatial allocation above. For example,
$\rho^{\mathrm w}=0.3$ installs 450 MW of wind in total, while
$\rho^{\mathrm e}=0.5$ installs 125 MW/500 MWh of battery storage in total, or
62.5 MW/250 MWh at each battery storage bus. Fig.~\ref{fig:network} shows the candidate
resource locations and the highlighted delivery interfaces.

\subsection{Sequence of Resource Additions}
\label{sec:cases}

Resources are added in the order in which they grew in ERCOT: wind first,
then solar, then battery storage \cite{ercot2018sog}. Adding battery storage
last also allows its charging and discharging to respond to the net load
profile of the complete wind and solar portfolio. Four case groups define
this cumulative sequence. \textbf{B1} is the thermal baseline,
$(\rho^{\mathrm w},\rho^{\mathrm s},\rho^{\mathrm e})=(0,0,0)$.
\textbf{B2} increases $\rho^{\mathrm w}$ from 0 to 1 with no solar or battery
storage. \textbf{B3} fixes $\rho^{\mathrm w}=1$ and increases
$\rho^{\mathrm s}$ from 0 to 1. \textbf{B4} fixes
$\rho^{\mathrm w}=\rho^{\mathrm s}=1$ and increases $\rho^{\mathrm e}$ from
0 to 1. In B2--B4, the build-out level takes the values
$\{0,0.1,0.2,\ldots,1\}$. Because adjacent case groups share an endpoint,
the sequence contains 31 distinct portfolio states.

The load, outage schedule, renewable availability, transmission network, and
operating assumptions are identical in every state. One build-out level changes
at a time, while unit commitment, dispatch, reserve provision, battery storage
operation, and line flows are reoptimized. Differences between adjacent states
therefore describe operating responses along the prescribed sequence. The
model is implemented in Pyomo (version 6.10.0) \cite{pyomo} with Python 3.12.10
and solved using Gurobi 12.0.3 \cite{gurobi}. Every instance reaches optimal
status at a relative mixed-integer programming gap tolerance of $10^{-9}$.

\section{Operating Model and Metrics}
\label{sec:model}

\subsection{Day-Ahead Unit Commitment With Network Constraints}
\label{sec:uc}

The operating model is a 24-hour day-ahead unit commitment formulation with
lossless DC power flow, branch thermal limits, joint scheduling of energy and
upward operating reserve, and chronological battery storage
\cite{shahidehpour,carrion}. Related day-ahead unit commitment formulations
have also been applied to operating studies of renewable power grids \cite{1,zhao2025tste}. The model
is formulated as a mixed-integer quadratic program with quadratic thermal
production costs, startup and shutdown costs, minimum up and down time
constraints, and ramp limits.

Let $\overline P^{\mathrm w}_b$ and $\overline P^{\mathrm s}_b$ denote the
candidate wind and solar nameplate capacities at bus $b$, and let
$\alpha^{\mathrm w}_{b,t},\alpha^{\mathrm s}_{b,t}\in[0,1]$ denote their
hourly availability factors. If $p^{\mathrm w}_{b,t}$ and
$p^{\mathrm s}_{b,t}$ are the dispatched wind and solar outputs, then
$0\le p^{\mathrm w}_{b,t}\le
\rho^{\mathrm w}\overline P^{\mathrm w}_b\alpha^{\mathrm w}_{b,t}$ and
$0\le p^{\mathrm s}_{b,t}\le
\rho^{\mathrm s}\overline P^{\mathrm s}_b\alpha^{\mathrm s}_{b,t}$.
The difference between available and dispatched renewable output is
curtailment.

Let $\mathcal B^{\mathrm e}\subseteq\mathcal B$ denote the set of battery storage
buses. At battery storage bus $b\in\mathcal B^{\mathrm e}$,
$\overline P^{\mathrm e}_b$ and $\overline E^{\mathrm e}_b$ denote the
candidate power and energy ratings. The installed ratings are
$\rho^{\mathrm e}\overline P^{\mathrm e}_b$ and
$\rho^{\mathrm e}\overline E^{\mathrm e}_b$. Let
$p^{\mathrm{ch}}_{b,t}$, $p^{\mathrm{dis}}_{b,t}$, and $e_{b,t}$ denote
charging power, discharging power, and stored energy. Charging and discharging
are each limited by the installed power rating, and a binary operating mode
prevents simultaneous charging and discharging. Stored energy evolves as
\begin{equation}
e_{b,t}=e_{b,t-1}
+\eta^{\mathrm{ch}}p^{\mathrm{ch}}_{b,t}\Delta t
-\frac{p^{\mathrm{dis}}_{b,t}}{\eta^{\mathrm{dis}}}\Delta t,
\label{eq:soc}
\end{equation}
where $\eta^{\mathrm{ch}}=\eta^{\mathrm{dis}}=\sqrt{0.86}$, giving a
round trip efficiency of 86\%. Stored energy is restricted to
$0.1\rho^{\mathrm e}\overline E^{\mathrm e}_b\le e_{b,t}\le
0.9\rho^{\mathrm e}\overline E^{\mathrm e}_b$. The initial and terminal
state of charge (SOC) are both fixed at 50\% of the installed energy rating:
\begin{equation}
e_{b,0}=e_{b,24}=0.5\rho^{\mathrm e}\overline E^{\mathrm e}_b.
\label{eq:soc_bc}
\end{equation}
Thus, each battery ends the day with the same stored energy as it starts,
and all energy it discharges is charged within the same day. The objective
includes a throughput based cycling cost of \$3/MWh, applied to both charging
and discharging energy.

For branch $l\in\mathcal L$,
$f_{l,t}$ denotes active power flow in interval $t$, and $\overline F_l$
denotes the branch thermal limit. The flow satisfies
$-\overline F_l\le f_{l,t}\le\overline F_l$. The voltage angle at bus 13
is fixed to zero in every interval. The modeled daily operating cost reported
below is the optimized 24-hour objective value under the stated cost and
penalty terms. After each mixed-integer solution, the binary variables are
fixed at their optimal values and the resulting convex quadratic program is
solved again; all dual variables reported in this paper are taken from this
fixed binary solution.

\subsection{Modeled Reserve Scarcity Pricing}
\label{sec:reserve}

The model co-optimizes energy and upward reserve \cite{isone2024cooptimization}.
Reserve scarcity is priced with the 2024 ERCOT ORDC \cite{ercot}. To fit the
test system, the reserve quantity axis of this curve is scaled by
$k=3{,}100/85{,}500=0.0363$, the ratio of test system peak demand to the
ERCOT reference peak. This scaling keeps reserve needs proportional to system
size: a reserve level that is the same fraction of peak demand receives the
same scarcity price in both systems. The price axis is unchanged, since ORDC
prices measure scarcity value per MWh, which does not depend on system size.
The modeled ORDC maximum is therefore \$5{,}000/MWh.

The approximation covers reserve quantities up to $R^{\max}=320$ MW and uses
16 segments indexed by $\mathcal S=\{1,\ldots,16\}$. Each segment has width
$\Delta R=20$ MW. Segment $s$ corresponds to the interval of available reserve
$[R^{\max}-s\Delta R,\,R^{\max}-(s-1)\Delta R]$ and has midpoint
$R_s^{\mathrm{mid}}=R^{\max}-(s-\tfrac{1}{2})\Delta R$. Its marginal
shortfall penalty $\kappa_s$ is the ERCOT ORDC value evaluated at the
corresponding reserve quantity on the ERCOT scale, $R_s^{\mathrm{mid}}/k$. Because
scarcity value increases as available reserve declines,
$\kappa_1\le\kappa_2\le\cdots\le\kappa_{16}$.

Let $\mathcal G$ denote the set of thermal units and
$\mathcal G^{\mathrm{qs}}\subseteq\mathcal G$ the set of quick start units.
For thermal unit $g$, $r_{g,t}$ denotes online upward reserve, $u_{g,t}$
denotes commitment status, and $\overline P_{g,t}$ denotes maximum available
output, which varies with the outage and derating schedule. For battery storage bus
$b\in\mathcal B^{\mathrm e}$, $r^{\mathrm e}_{b,t}$ denotes upward reserve
from battery storage, limited by the discharging power headroom and by the energy
stored above the minimum SOC. The variable $q_{s,t}$ denotes reserve shortfall
assigned to segment $s$, with $0\le q_{s,t}\le\Delta R$. The hourly reserve
balance is
\begin{equation}
\begin{aligned}
&\sum_{g\in\mathcal G}r_{g,t}
+\sum_{g\in\mathcal G^{\mathrm{qs}}}\overline P_{g,t}(1-u_{g,t}) \\
&\qquad
+\sum_{b\in\mathcal B^{\mathrm e}}r^{\mathrm e}_{b,t}
+\sum_{s\in\mathcal S}q_{s,t}
\ge R^{\max}.
\end{aligned}
\label{eq:res}
\end{equation}
The first three terms in \eqref{eq:res} represent upward reserve from online
thermal units, offline quick start units, and battery storage, respectively. The final
term represents reserve shortfall relative to $R^{\max}$. The objective
includes
$\sum_{t\in\mathcal T}\sum_{s\in\mathcal S}\kappa_s q_{s,t}\Delta t$, so
lower cost segments are used before higher cost segments as available reserve
declines.

The dual variable $\pi_t^{\mathrm r}$ associated with \eqref{eq:res} in the
fixed binary solution is the hourly modeled reserve scarcity price. It measures
the marginal increase in operating cost from a 1 MW increase in $R^{\max}$ in
interval $t$. Because $\Delta t=1~\mathrm{h}$, the dual is numerically
reported in \$/MWh; when segment $s$ is marginal,
$\pi_t^{\mathrm r}=\kappa_s$.

The value of lost load, $V^{\mathrm{ll}}=\$9{,}000/\mathrm{MWh}$, exceeds the
maximum reserve shortfall penalty. When available capacity cannot
simultaneously serve demand and provide the full reserve quantity, the model
therefore accepts reserve shortfall before leaving firm load unserved.

\subsection{Prices and Reported Metrics}
\label{subsec:lmp}

The dual of the nodal power balance in the fixed binary solution is the
locational marginal price (LMP) $\lambda_{b,t}$. These prices are conditional
on the optimized commitment and battery storage modes. The lossless network has no
marginal loss component.

Bus 13 is used as the reporting reference bus. Define the energy component
as $\lambda_t^{\mathrm{en}}=\lambda_{13,t}-\pi_t^{\mathrm r}$ and
the congestion component as
$\lambda_{b,t}^{\mathrm c}=\lambda_{b,t}-\lambda_{13,t}$. The nodal price then
satisfies
\begin{equation}
\lambda_{b,t}
=\underbrace{\lambda_t^{\mathrm{en}}}_{\text{energy}}
+\underbrace{\pi_t^{\mathrm r}}_{\text{modeled reserve scarcity}}
+\underbrace{\lambda_{b,t}^{\mathrm c}}_{\text{congestion}}.
\label{eq:decomp}
\end{equation}
Equation~\eqref{eq:decomp} is an accounting identity under these definitions.
The energy and congestion components depend on the selected reference bus, whereas their sum with the modeled reserve scarcity component recovers the
nodal LMP. The components are reported as load weighted daily averages using
\eqref{eq:weighted_average}.

Gross load payment is
$L^{\mathrm{gross}}=
\sum_{b\in\mathcal B}\sum_{t\in\mathcal T}
\lambda_{b,t}D_{b,t}\Delta t$.
This quantity is reported only for portfolio states with
$E^{\mathrm{sh}}=0$, for which all demand is served.

Let $\mu^{+}_{l,t}\geq 0$ and $\mu^{-}_{l,t}\geq 0$ be the duals of the
upper and lower branch flow limits, respectively, and define
$\widetilde{\mu}_{l,t}=\mu^{+}_{l,t}-\mu^{-}_{l,t}$. Daily congestion
rent is
\begin{equation}
M^{\mathrm{cong}}
=
\sum_{l\in\mathcal L}\sum_{t\in\mathcal T}
\widetilde{\mu}_{l,t}f_{l,t}\Delta t
\geq 0.
\label{eq:congestion_rent}
\end{equation}

For renewable type $x\in\{\mathrm w,\mathrm s\}$ with positive dispatched
energy over the study day, the daily capture price at bus $b$ is
\begin{equation}
\psi^x_b
=
\frac{\displaystyle
\sum_{t\in\mathcal T}\lambda_{b,t}p^x_{b,t}\Delta t}
{\displaystyle
\sum_{t\in\mathcal T}p^x_{b,t}\Delta t}.
\label{eq:capture_price}
\end{equation}
For $b\in\mathcal B^{\mathrm e}$ and $\rho^{\mathrm e}>0$, normalized daily net
energy market revenue is
\begin{equation}
\nu^{\mathrm e}_b
=
\frac{\displaystyle
\sum_{t\in\mathcal T}\lambda_{b,t}
\left(p^{\mathrm{dis}}_{b,t}-p^{\mathrm{ch}}_{b,t}\right)\Delta t}
{\displaystyle
\rho^{\mathrm e}\overline P^{\mathrm e}_b}.
\label{eq:storage_revenue}
\end{equation}
Capture prices exclude production incentive revenue. Equation
\eqref{eq:storage_revenue} includes energy market settlements but excludes
reserve payments, cycling costs, and investment costs. Because the initial and
terminal SOC are equal by \eqref{eq:soc_bc}, the 24-hour settlement is
evaluated without net depletion of stored energy.

\section{Results}
\label{sec:results}

\begin{figure}[!t]
\centering
\includegraphics[width=\columnwidth]{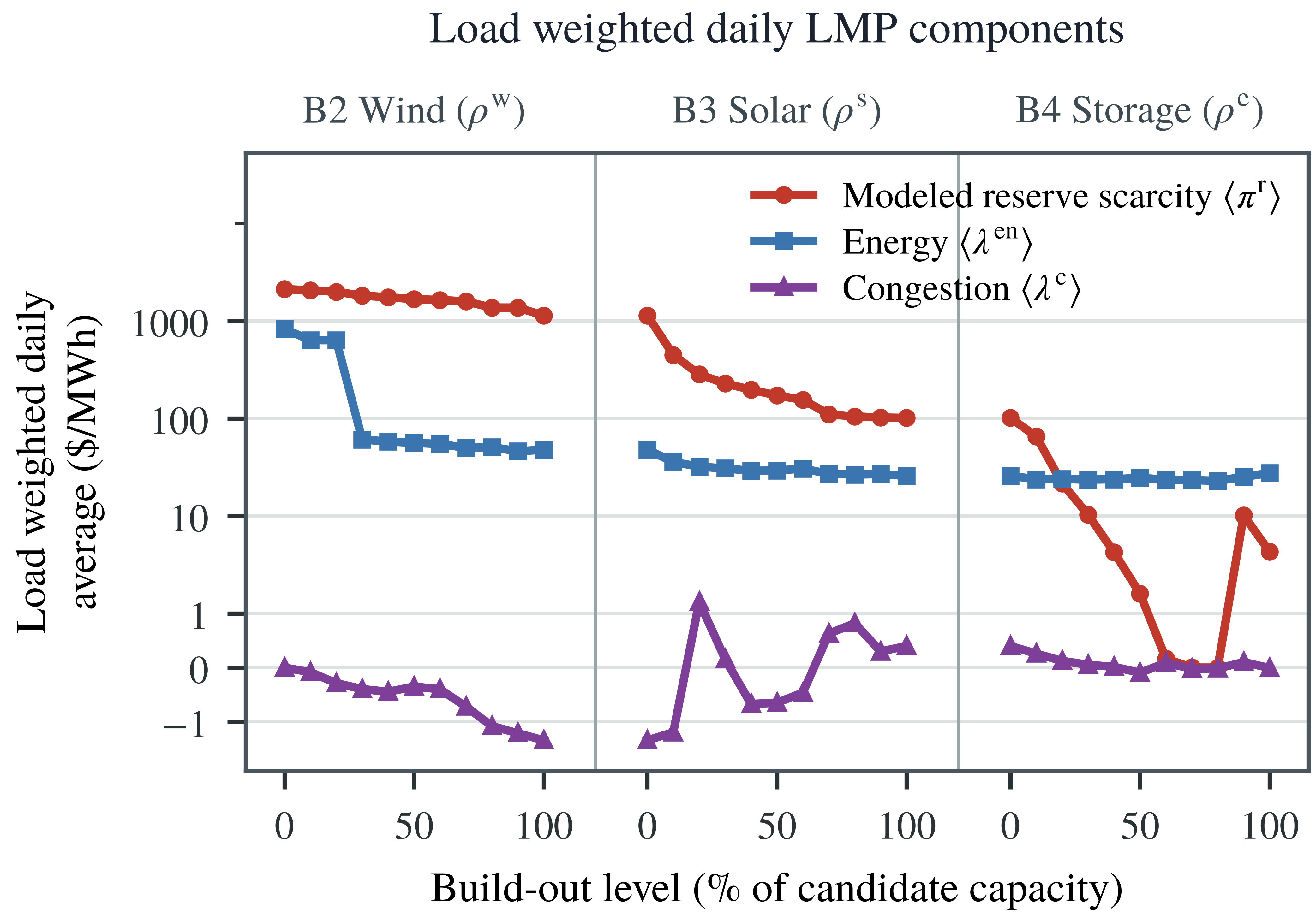}
\vspace{-5mm}
\caption{Load weighted daily averages of the three LMP components in
\eqref{eq:decomp} across the 31 portfolio states, with B1 at the left end of
B2. The symmetric logarithmic axis, linear between $-\$1$/MWh and \$1/MWh,
displays values that are both positive and negative and span several orders
of magnitude.}
\label{fig:decomp}
\end{figure}

Table~\ref{tab:summary} reports selected systemwide outcomes,
Fig.~\ref{fig:decomp} shows the load weighted LMP components across all 31
states, and Table~\ref{tab:value} reports selected locational
outcomes for the portfolio with all candidate resources installed. Along the
prescribed sequence, unserved energy is eliminated and modeled reserve scarcity
falls substantially, while transmission congestion, curtailment, and
locational differences remain.

\begin{table*}[!t]
\caption{Selected Systemwide Operating Outcomes Along the Prescribed Resource Sequence}
\label{tab:summary}
\centering
\footnotesize
\setlength{\tabcolsep}{5pt}
\begin{tabular}{@{}lp{3.6cm}rrrp{4.9cm}@{}}
\toprule
\shortstack[l]{Case\\group}
& Portfolio state
& \shortstack{Unserved\\energy\\(MWh)}
& \shortstack{Wind\\curtailment\\(MWh)}
& \shortstack{Load weighted daily average\\modeled reserve scarcity\\price (\$/MWh)}
& Main observation \\
\midrule
B1
& Thermal baseline
& 106
& 0
& 2{,}116.3
& Unserved energy is present. \\
B2 ($\rho^{\mathrm w}=0.3$)
& 450 MW wind
& 0
& 0
& 1{,}817.3
& All demand is served; modeled reserve scarcity remains. \\
B2 ($\rho^{\mathrm w}=1$)
& 1{,}500 MW wind
& 0
& 631
& 1{,}136.2
& Evening modeled reserve scarcity and wind curtailment remain. \\
B3 ($\rho^{\mathrm s}=1$)
& $+$ 1{,}800 MW solar
& 0
& 631
& 101.2
& Positive modeled reserve scarcity prices are confined to three evening hours. \\
B4 ($\rho^{\mathrm e}=1$)
& $+$ 250 MW/1{,}000 MWh\newline battery storage
& 0
& 306
& 4.3
& Wind curtailment falls further; transmission congestion remains. \\
\bottomrule
\end{tabular}

\vspace{2pt}
\parbox{0.98\textwidth}{\footnotesize
Resource additions are cumulative. All battery storage cases enforce equal 50\%
initial and terminal SOC.}
\vspace{-2mm}
\end{table*}

\subsection{Wind Eliminates Unserved Energy During the Peak Hours}

The thermal baseline has 106 MWh of unserved energy from HE15 through HE18,
including 42.0 MW of unserved load in each of HE16 and HE17. With 450 MW of the
candidate wind capacity installed, wind output is 43.2 MW in both peak hours
and ranges from 31.7 MW to 43.2 MW over the four affected intervals. This is the first portfolio state with $E^{\mathrm{sh}} = 0$.

Modeled reserve scarcity remains after all demand is served. With all
1{,}500 MW of candidate wind installed, wind output in HE20 is 66.6 MW, or
4.4\% of installed wind capacity. The hourly modeled reserve scarcity price reaches the
modeled ORDC maximum of \$5{,}000/MWh in HE18 and HE19 and is
\$3{,}315/MWh in HE20. Wind additions reduce the load weighted daily average
modeled reserve scarcity price, but the evening decline in wind output leaves limited
upward flexibility. Eliminating the energy shortfall at the peak therefore does
not eliminate modeled reserve scarcity later in the day.

\vspace{-3pt}
\subsection{Solar and Battery Storage Reduce Evening Modeled Reserve Scarcity}

\begin{figure}[!t]
\centering
\includegraphics[width=\columnwidth]{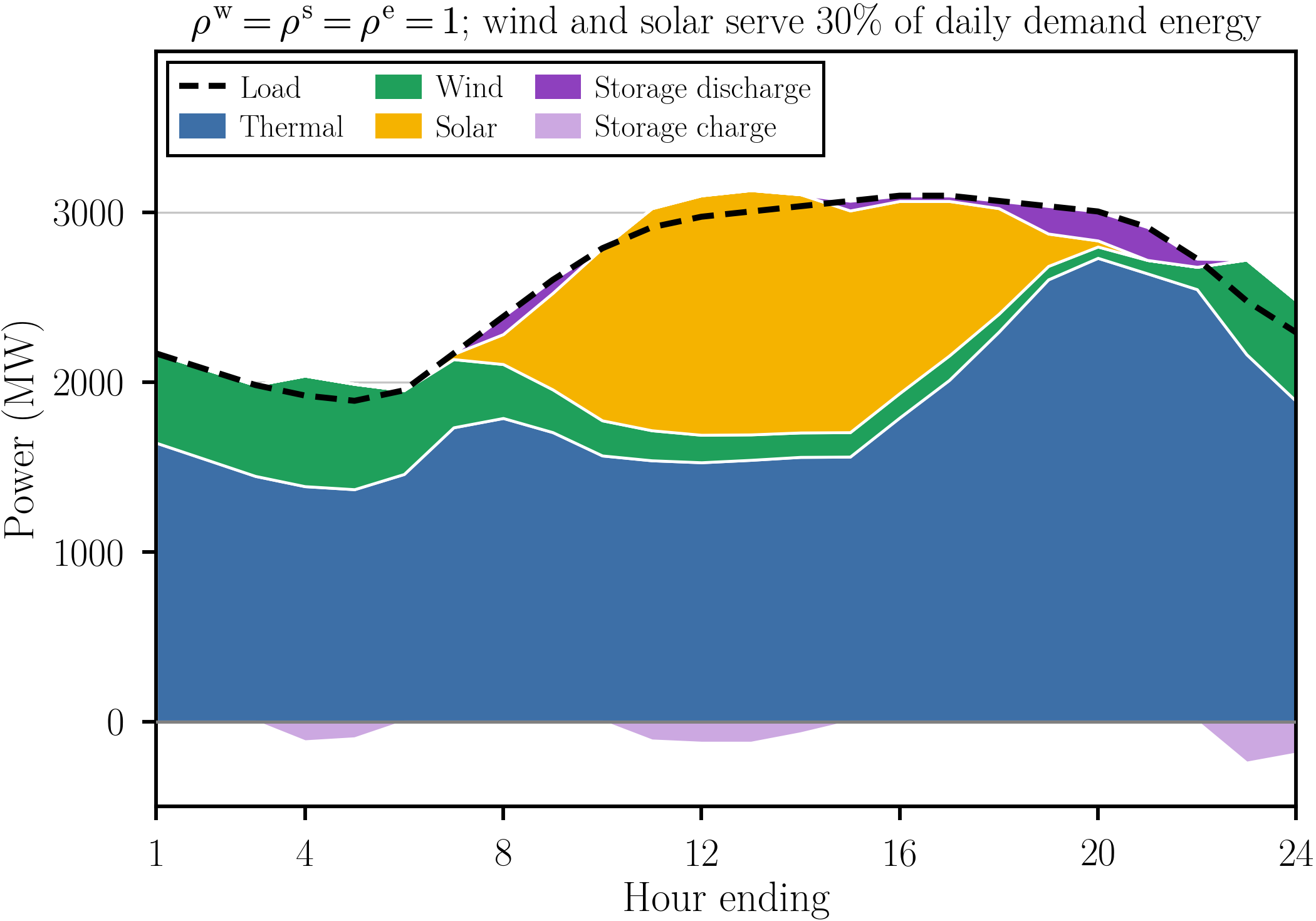}
\caption{Hourly operation of the portfolio with all candidate resources
installed ($\rho^{\mathrm w}=\rho^{\mathrm s}=\rho^{\mathrm e}=1$).
}
\label{fig:dispatch}
\end{figure}

With all candidate wind installed, the modeled reserve scarcity price is
positive in ten hours. Adding all candidate solar reduces its load weighted
daily average by 91\%, from \$1{,}136.2/MWh to \$101.2/MWh, and confines
positive values to HE19--HE21. The highest remaining value is \$1{,}904/MWh in
HE20, when solar output has fallen to 36.0 MW. For this case, renewable
availability during HE19--HE21 is more relevant to the remaining modeled reserve scarcity than
total renewable energy over the full day.

Installing all 250 MW/1{,}000 MWh of candidate battery storage reduces the
load weighted daily average modeled reserve scarcity price from \$101.2/MWh to
\$4.3/MWh, a decrease of about 96\% relative to the state with all candidate
solar and no battery storage. Wind curtailment falls from 631 MWh to 306 MWh. The
modeled daily operating cost of the state with full battery storage is \$641{,}328. The
battery storage addition therefore reduces both evening modeled reserve scarcity and
wind curtailment.

Fig.~\ref{fig:dispatch} shows the hourly operation behind these outcomes.
Solar output falls from 622.8 MW in HE18 to 190.8 MW in HE19 and 36.0 MW in
HE20 while demand is still above 3{,}000 MW, thermal generation reaches its
daily maximum of 2{,}732.6 MW in HE20, and battery storage discharges 162.3 MW,
171.8 MW, and 193.8 MW in HE19--HE21. Charging occurs in HE4--HE5,
HE11--HE14, and HE23--HE24; the last interval coincides with recovering wind
output and with the terminal SOC condition. Positive
modeled reserve scarcity prices in this state occur only in HE20 and HE21, and
these two hours account for the \$4.3/MWh load weighted daily average in
Table~\ref{tab:summary}.

However, Fig.~\ref{fig:decomp} also shows that lower operating cost does not
always mean lower modeled reserve scarcity prices: the price rebounds at high
battery storage build-out levels. This rebound is a unit commitment
discontinuity. Because commitment decisions are binary, a small increase in
battery storage can switch a unit off and change the price in a single step.
Supplementary runs at finer build-out increments place this step between
$\rho^{\mathrm e}=0.85$ and $0.86$. At this step, the added battery storage
makes it economical to keep a 155 MW coal unit offline, which removes its
reserve contribution. Two quick start units are committed at the same time
and no longer count as offline reserve, and greater battery discharge reduces
battery reserve headroom. With less reserve available, the load weighted
daily average modeled reserve scarcity price rises sharply from near zero,
even though operating cost falls and all demand remains served. Further
battery storage additions lower the price again.

\vspace{-4pt}
\subsection{Transmission Congestion and Locational Outcomes}
\vspace{-2pt}
Transmission congestion persists after modeled reserve scarcity is
substantially reduced. Across the solar and battery storage additions, gross load
payment falls from \$74.1 million with all candidate wind installed to
\$8.0 million after the full solar addition and \$2.0 million after the
full battery storage addition. Over the same stages, daily congestion rent
remains between \$478{,}000 and \$687{,}000.

The load weighted congestion component $\langle\lambda^{\mathrm c}\rangle$ in
Fig.~\ref{fig:decomp} stays within $\pm$\$1.6/MWh along the entire sequence,
while daily congestion rent remains in the hundreds of thousands of dollars.
The two quantities describe different aspects of congestion. Positive and
negative nodal deviations from the reference bus offset each other in the
load weighted average, so $\langle\lambda^{\mathrm c}\rangle$ measures the
effect of congestion on the systemwide average price, which is small here.

With all candidate resources installed, all 306 MWh of wind curtailment occurs
at bus 21, while wind at buses 22 and 23 is not curtailed. The daily wind
capture price defined in \eqref{eq:capture_price} is
$-\$7.3/\mathrm{MWh}$ at bus 21 and \$24.5/MWh at bus 23, a difference of
\$31.8/MWh. In hours with high wind availability, exports from bus 21 are
limited by the interface to bus 15 (Fig.~\ref{fig:network}), and the LMP at
bus 21 is set by the negative wind offer cost. Much of the wind output at
bus 21 is produced in these hours, so its capture price, which weights LMPs
by output, falls below zero, while LMPs at bus 23 remain positive in the same
hours. The concentration of curtailment and the separation in capture prices
show that wind projects with the same technology and offer assumptions can
face substantially different outcomes at different network locations.

The identical 125 MW/500 MWh battery storage plants also have different energy market
outcomes. Their normalized daily net energy market revenues, defined in
\eqref{eq:storage_revenue}, are
\$201/(MW$\cdot$day) at bus 19 and \$121/(MW$\cdot$day) at bus 21. The revenue difference
reflects their nodal price profiles and the resulting charging and discharging
schedules.

\begin{table}[!t]
\caption{Selected Locational Operating and Energy Market Outcomes With All Candidate Resources Installed}
\label{tab:value}
\centering
\footnotesize
\setlength{\tabcolsep}{5pt}
\textit{Wind outcomes at selected buses}\\[2pt]
\begin{tabular}{lrr}
\toprule
Bus & \shortstack{Capture price\\(\$/MWh)}
& \shortstack{Wind curtailment\\(MWh)} \\
\midrule
21 & $-7.3$ & 306 \\
23 & 24.5 & 0 \\
\bottomrule
\end{tabular}

\vspace{4pt}
\textit{Battery storage energy market outcomes}\\[2pt]
\begin{tabular}{lr}
\toprule
Bus & \shortstack{Normalized net\\energy market revenue\\
$\bigl(\$/(\mathrm{MW}\cdot\mathrm{day})\bigr)$} \\
\midrule
19 & 201 \\
21 & 121 \\
\bottomrule
\end{tabular}
\\[2pt]
\parbox{0.96\columnwidth}{\footnotesize
Wind at bus 22 has zero curtailment and is omitted.}
\vspace{-2mm}
\end{table}

\section{Conclusion}
\label{sec:conclusion}

On the constructed stressed day, the thermal baseline has 106 MWh of unserved
energy, and 450 MW of candidate wind is the first portfolio state that serves
all demand. Modeled reserve scarcity nevertheless remains with all candidate
wind installed. Adding all candidate solar reduces the load weighted daily
average modeled reserve scarcity price from \$1{,}136.2/MWh to \$101.2/MWh. With all
candidate battery storage installed, the average is \$4.3/MWh, wind curtailment is
306 MWh, and the modeled daily operating cost is \$641{,}328.

Transmission congestion continues to shape locational outcomes after
systemwide modeled reserve scarcity is substantially reduced. All residual wind
curtailment occurs at bus 21; wind capture prices at buses 21 and 23 differ by
\$31.8/MWh; and the identical battery storage plants yield normalized net energy market
revenues of \$201/(MW$\cdot$day) at bus 19 and \$121/(MW$\cdot$day) at bus 21, respectively. These results show
that unserved energy, modeled reserve scarcity, and the locational effects of
transmission congestion measure different operating conditions and should be
reported separately but interpreted together when evaluating sequential wind,
solar, and battery storage additions.

\section*{Acknowledgment}

This work was supported by the National Academy of Engineering Grainger Foundation Frontiers of Engineering Grant. This work was completed in part with resources provided by the Research Computing Data Core at the University of Houston.

\bibliographystyle{IEEEtran}
\bibliography{references}

\end{document}